# Effect of Native Defects on Optical Properties of $In_xGa_{1-x}N$ Alloys

S. X. Li and E.E. Haller,
Materials Sciences Division, Lawrence Berkeley National Laboratory, and Department of Materials Science and Engineering, University of California, Berkeley, California 94720
K. M. Yu, W. Walukiewicz, J. W. Ager III, J. Wu, and W. Shan,
Materials Sciences Division, Lawrence Berkeley National Laboratory, Berkeley, California 94720
Hai Lu and William J. Schaff
Department of Electrical and Computer Engineering, Cornell University, Ithaca, New York 14853

Abstract:

The energy position of the optical absorption edge and the free carrier populations in $In_xGa_{1-x}N$ ternary alloys can be controlled using high energy $^4He^+$ irradiation. The blue shift of the absorption edge after irradiation in In-rich material ($x > 0.34$) is attributed to the band-filling effect (Burstein-Moss shift) due to the native donors introduced by the irradiation. In Ga-rich material, optical absorption measurements show that the irradiation-introduced native defects are inside the bandgap, where they are incorporated as acceptors. The observed irradiation-produced changes in the optical absorption edge and the carrier populations in $In_xGa_{1-x}N$ are in excellent agreement with the predictions of the amphoteric defect model.

PACS numbers: 78.66.Fd, 78.40.Fy

Electronic Mail: w_walukiewicz@lbl.gov

Since the discovery of its narrow (0.7 eV) bandgap [1, 2], InN has attracted intense research interest. Interestingly, the actual value of its direct bandgap has been the subject of some controversy [3-8]. In most cases, measurements of apparent band gaps larger than 0.7 eV can be attributed to the significant Burstein-Moss shift (conduction band filling) that occurs at high electron concentrations [9]. Because it has the highest electron affinity (~5.8 eV) in known semiconductors, InN is extremely susceptible to n-type doping, either by native donors or by impurities such as oxygen. The relatively small electron effective mass and high energy position of other conduction band valleys (e.g. L and X) make the conduction band-filling effect particularly strong in InN. For example, for an electron concentration in the mid-$10^{20}$cm$^{-3}$ range, the absorption edge in InN shifts to ~1.7 eV [9].

We have shown previously [10] that native donors and/or acceptors introduced through energetic particle irradiation can be used to precisely control the free electron concentration in $In_{1-x}Ga_xN$ alloys. Here, we show that the defects have a profound effect on the optical absorption properties as well. Our results suggest that energetic particle irradiation provides a simple and convenient way of altering the doping and optical absorption characteristics of $In_xGa_{1-x}N$ alloys in a controllable manner. This technique, which is applicable to all group-III nitride alloys and could be performed potentially with a focused ion beam, could open up novel applications of these alloys.

Epitaxial InN and $In_xGa_{1-x}N$ thin films (310-2700 nm thick) used in this study were grown on c-sapphire substrates by molecular beam epitaxy (MBE) with a GaN (~200 nm thick) buffer layer [11]. The initial free electron concentrations in these samples ranged from the low $10^{18}$ cm$^{-3}$ to low $10^{17}$ cm$^{-3}$ and the mobility ranged from 7 cm$^2$/V·s ($x$ = 0.24) to above 1500 cm$^2$/Vs ($x$ = 1). A GaN sample (3 μm thick) grown by metal-organic chemical vapor deposition (MOCVD) with an initial electron concentration of $7.74\times10^{17}$ cm$^{-3}$ and a mobility of 189 cm$^2$/V·s was also used in this study.

The samples were irradiated with 2 MeV $^4He^+$ ions generated by a Van de Graaff accelerator, with fluences between $1.12\times10^{14}$ and $2.68\times10^{16}$ cm$^{-2}$. In all cases, the particle penetration depth exceeded the film thickness, assuring homogeneous generation of defects in the film. Ion channeling spectroscopy showed that the minimum yield $\chi$ increased from 0.04 in an as-grown InN sample to only 0.11 after $^4He^+$ irradiation with a dose of $1.8\times10^{16}$ cm$^{-2}$, indicating that the InN film remains single crystalline. Cross-sectional transmission electron microscopy of an InN sample subjected to the heaviest dose of irradiation showed that no additional extended defects were formed, indicating that point defects are responsible for the observed changes in electrical and optical properties of the irradiated samples. The optical absorption measurements were performed at room temperature using a CARY-2390 NIR-VIS-UV spectrophotometer. Free electron concentration and mobility were measured at room temperature using a home-built Hall effect system with a magnetic field of 3000 Gauss.

Figure 1 shows the evolution of the optical absorption spectra of InN and $In_{0.4}Ga_{0.6}N$ with increasing irradiation dose. In both samples, the absorption coefficient is on the order of $5x10^4$ cm$^{-1}$ at 0.5 eV above the absorption onset, which is typical for direct bandgap semiconductors. With increasing irradiation dose the absorption edges show a blue shift. More specifically, the absorption edge shifts to higher energy, while the baseline and the slope above the absorption edge remain unchanged. The shift is composition-dependent, with smaller shifts found in the samples with higher Ga content. From as-grown to the highest dose of $^4He^+$ irradiation of $2.68\times10^{16}$ cm$^{-2}$ the absorption edge shifted by 1.05eV in InN, by 0.71 eV in $In_{0.7}Ga_{0.3}N$ (data not shown), but by only 0.15 eV in $In_{0.4}Ga_{0.6}N$. The shift is also observed to saturate with radiation dose. For example, as seen in Fig. 1 (a) for InN, the blue shift slows down as the irradiation dose increases and eventually becomes insensitive to further irradiation at a sufficiently high dose (typically > $10^{16}$ cm$^{-2}$). The behavior of the more Ga-rich ($x < 0.34$) material is different. As illustrated in Fig. 2, irradiation of GaN does not affect the fundamental absorption edge energy at ~3.4 eV but rather produces a new sub-bandgap absorption feature at ~2.7 eV. Both the strength and the linewidth of the absorption peak increase with increasing irradiation dose. Clearly unfilled or partially-filled defect states are formed inside the bandgap of GaN as a result of the irradiation.

Changes in the electronic properties of the irradiated $In_xGa_{1-x}N$ samples, the details of which have been reported separately [10], cause the changes in the optical properties. Despite having similar initial free electron concentrations, the $In_xGa_{1-x}N$ samples with different compositions behave very differently after irradiation. In $In_xGa_{1-x}N$ with $x > 0.34$ the electron concentration increases with increasing $^4He^+$ irradiation dose and eventually saturates at high (> $10^{16}$ $cm^{-2}$) doses. On the contrary, in Ga-rich $In_xGa_{1-x}N$ the free electron concentration decreases with increasing irradiation dose. In the case of GaN, the electron concentration decreased so quickly in the fluence range used that the sample became too resistive to be measured. Figure 3 shows the free electron saturation concentration of $In_xGa_{1-x}N$ at high dose of irradiation as a function of Ga fraction [10]. The saturation concentration decreases from $4.1\times10^{20}$ $cm^{-3}$ in InN to $6.5\times10^{14}$ $cm^{-3}$ in $In_{0.24}Ga_{0.76}N$.

Both the optical and the electrical properties can be understood using the amphoteric defect model (ADM) [12, 13]. The model predicts that in all semiconductors the native defects that are highly localized in nature have a common energy level, which is termed the Fermi-stabilization Energy ($E_{FS}$) and is located ~4.9 eV below the vacuum level. In the inset of Figure 3, the position of the $E_{FS}$ is shown together with the composition dependence of the valence and conduction band edges (VBE and CBE) of $In_xGa_{1-x}N$. $E_{FS}$ is located ~0.9 eV above the CBE of InN but ~0.7 eV below the CBE of GaN (2.7 eV above the VBE). According to the ADM either donor- or acceptor-like native point defects are formed depending on the relative position of the Fermi level ($E_F$) to the $E_{FS}$. In InN the $E_{FS}$ is above the CBE; as a result, donor-like irradiation-induced point defects form. Therefore, as the irradiation dose increases, the electron concentration increases until $E_F$ approaches $E_{FS}$. At this point both donor and acceptor-like defects are formed at similar rates and compensate each other leading to stabilization of $E_F$ and saturation of the electron concentration. As a result, a large increase and then a saturation in the Burstein-Moss shift of the optical absorption edge is predicted; this is in fact observed in Fig. 1(a).

With increasing Ga content in $In_xGa_{1-x}N$ alloys, the CBE moves towards $E_{FS}$, resulting in smaller electron saturation concentrations and absorption edge shifts. For $x <$ 0.34 the CBE moves above $E_{FS}$ and acceptors become the dominant irradiation-induced defects in n-type samples. For Ga-rich material, the acceptor defects compensate n-type conductivity until the Fermi energy stabilizes at $E_{FS}$. In the case of GaN, where the compensation is the most effective, the sample eventually became semi-insulating after irradiation. As expected, the decreasing electron concentration in Ga-rich $In_{1-x}Ga_xN$ does not have a significant effect on the absorption edge. However, the formation of the native defect states, which are predicted by the ADM to occur at 2.7 eV above the VBE in GaN (inset of Fig. 3), is observed clearly by optical absorption in Fig. 2.

To demonstrate the agreement between the optical and electronic properties of irradiated InN, the absorption spectra are numerically analyzed to obtain the Fermi energy. To account for broadening effects a Gaussian function was convoluted with the energy dependent absorption coefficient for a direct gap,

$$\alpha(E) = \frac{1}{\Delta\sqrt{\pi}} \int_{-\infty}^{\infty} \alpha_0(E') \left[1 + \exp\left(\frac{\mathrm{E_F} - E'}{kT}\right)\right]^{-1} \exp\left[-\left(\frac{E' - E}{\Delta}\right)^2\right] dE', \qquad (1)$$

where $\alpha_0(E')$ is the ideal absorption of InN and $\Delta$ is the Gaussian broadening parameter. The best fits, which are plotted in Fig. 1(a) as solid lines, are obtained by adjusting $\Delta$ and the Fermi energy ($E_F$). The corresponding $E_F$ values are labeled by arrows; the $\Delta$ values do not vary significantly and lie consistently between 0.21 - 0.23 eV. The $E_F$ values of InN and $In_xGa_{1-x}N$ alloys derived from absorption spectra show excellent agreement with the electron concentrations measured by room-temperature Hall effect, from a different set of irradiated samples. The consistency and repeatability suggest that the irradiation is a dependable method to control the doping and optical properties of $In_xGa_{1-x}N$ alloys.

In principle, the same irradiation effect, which is ultimately the displacement of lattice atoms, can be achieved in all group III-nitride alloys by using an equivalent dose of any other energetic particles such as electrons, protons, and other ions. By focusing the particle beam, one can alter the doping or optical properties of group III-nitride alloys with desired patterns down to the nanometer scale. This offers a range of possible applications of this technique for fabrication of highly conducting spatially confined structures.

In conclusion, we have shown that optical absorption properties of $In_xGa_{1-x}N$ alloys can be controlled by high energy particle irradiation. As predicted by the amphoteric defect model, native point defects introduced by irradiation are incorporated as donors in In-rich $In_xGa_{1-x}N$ ($x > 0.34$) but as acceptors in Ga-rich $In_xGa_{1-x}N$ ($x < 0.34$). The native donors increase the electron concentration and cause a blue shift of the absorption edge in In-rich $In_{1-x}Ga_xN$ while the native acceptors lower the electron concentration and form defect states in side the bandgap of Ga-rich $In_xGa_{1-x}N$, as observed by optical absorption.

We thank Milton Yeh of Blue Photonics Inc. for providing the GaN samples and Z. Liliental-Weber for the TEM results. This work is supported by the Director, Office of Science, Office of Basic Energy Sciences, Division of Materials Sciences and Engineering, of the U.S. Department of Energy under Contract No. DE-AC03-76SF00098. The work at Cornell University is supported by ONR under contract NO. N000149910936.

**Figure caption:**

Fig. 1: The evolution of InN and $In_{0.4}Ga_{0.6}N$ absorption spectra with increasing doses of 2 MeV $^4He^+$ irradiation. In Fig. 1 (a), the experimental results of InN are shown as data points and the numerical fit of the optical absorption spectra (see text) are shown as solid lines. The Fermi energies obtained from the fitting are marked by the arrows.

Fig. 2: The absorption spectra of GaN with increasing doses of 2 MeV $^4He^+$ irradiation.

Fig. 3: The saturation concentration of $In_xGa_{1-x}N$ ($1 \geq x \geq 0.24$) as a function of Ga fraction [10]. The data points are experimental results while the solid line is a numerical fit. In the inset, the band offset diagram of $In_{1-x}Ga_xN$ is shown with the energy level of $E_{FS}$ marked.

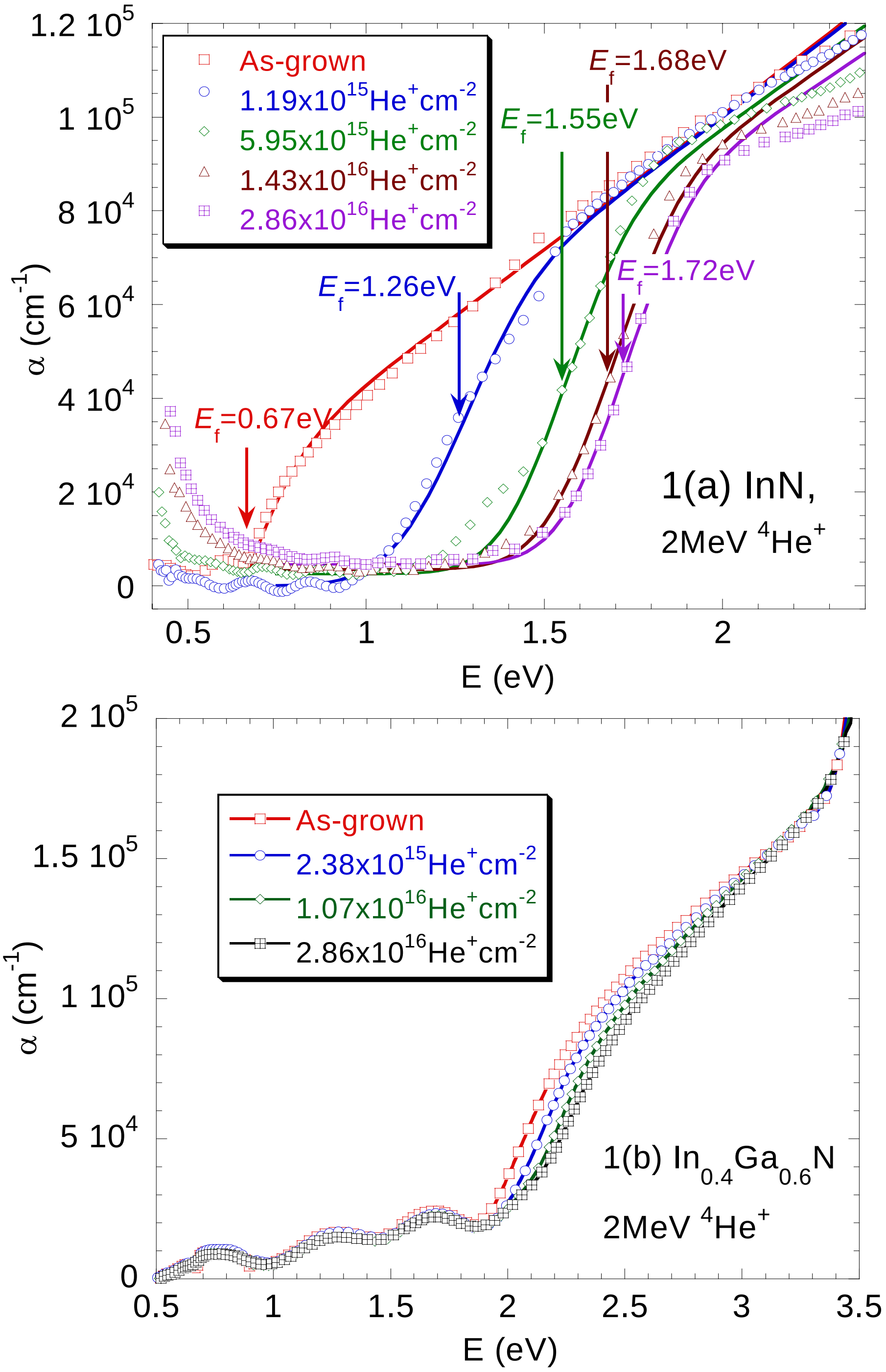

Fig. 1 of 3, the following 2 figures are the individual Fig.1 (a) and (b)

Li *et al*.

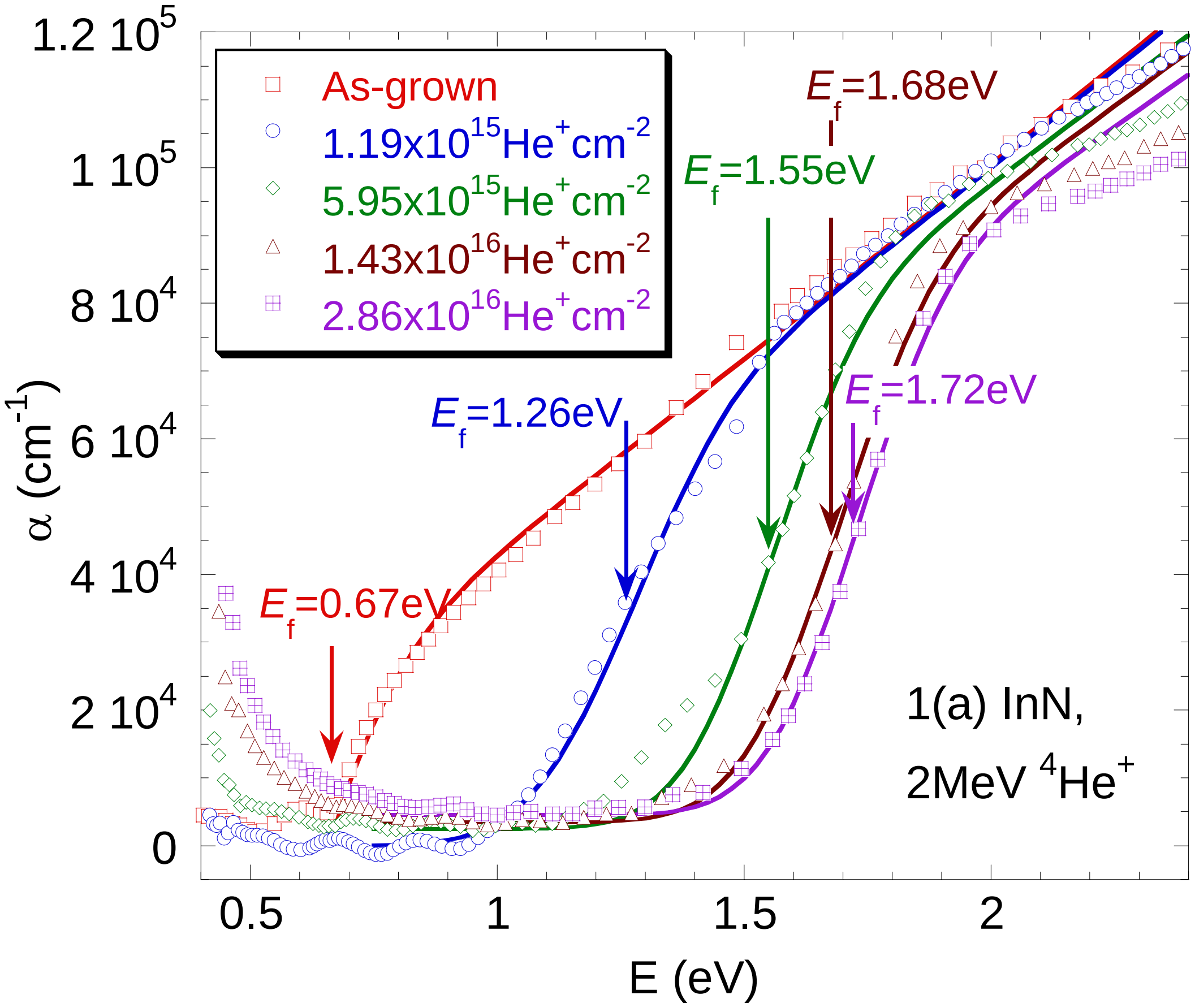

Fig. 1(a)

Li *et al.*

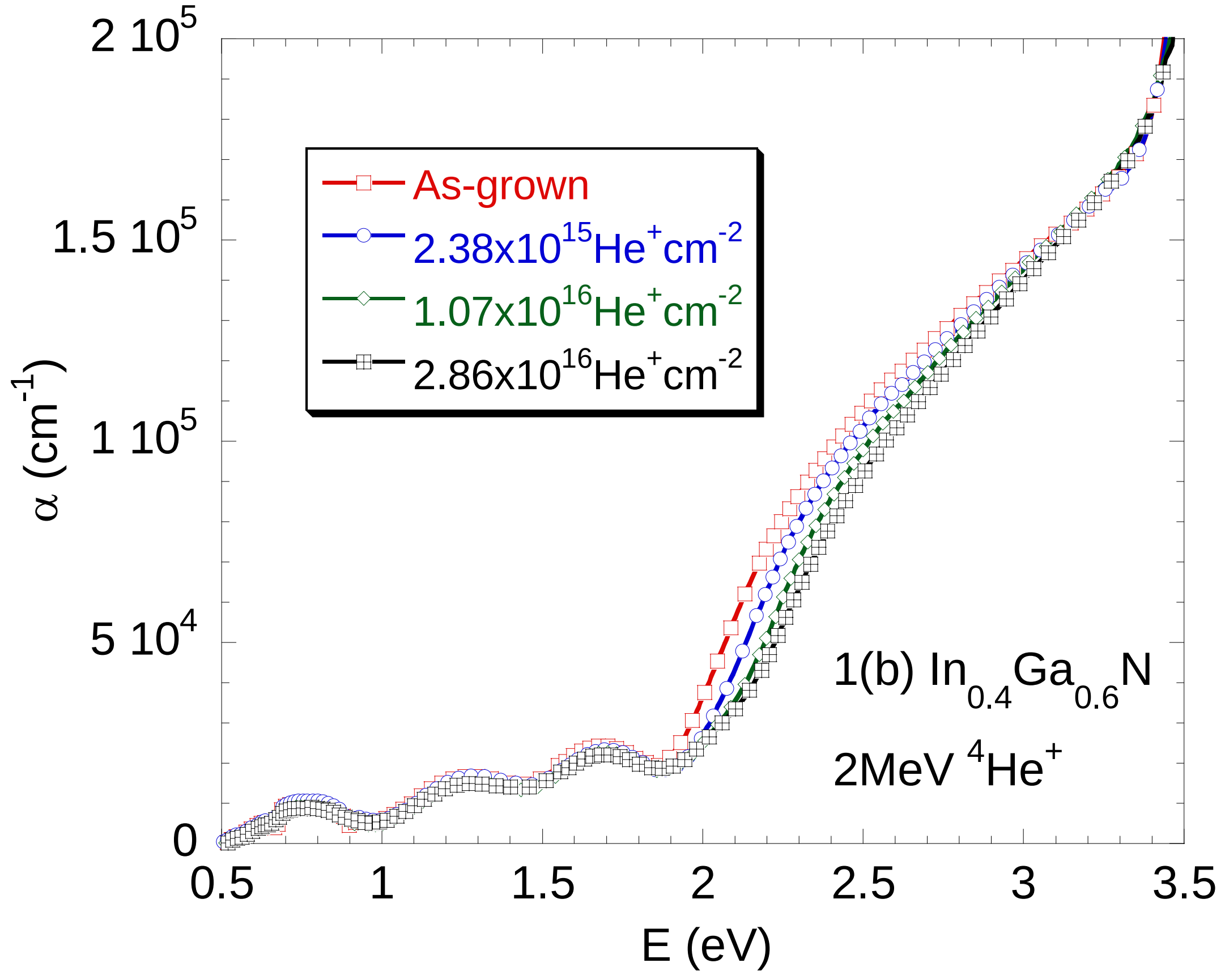

Fig. 1(b)

Li *et al*.

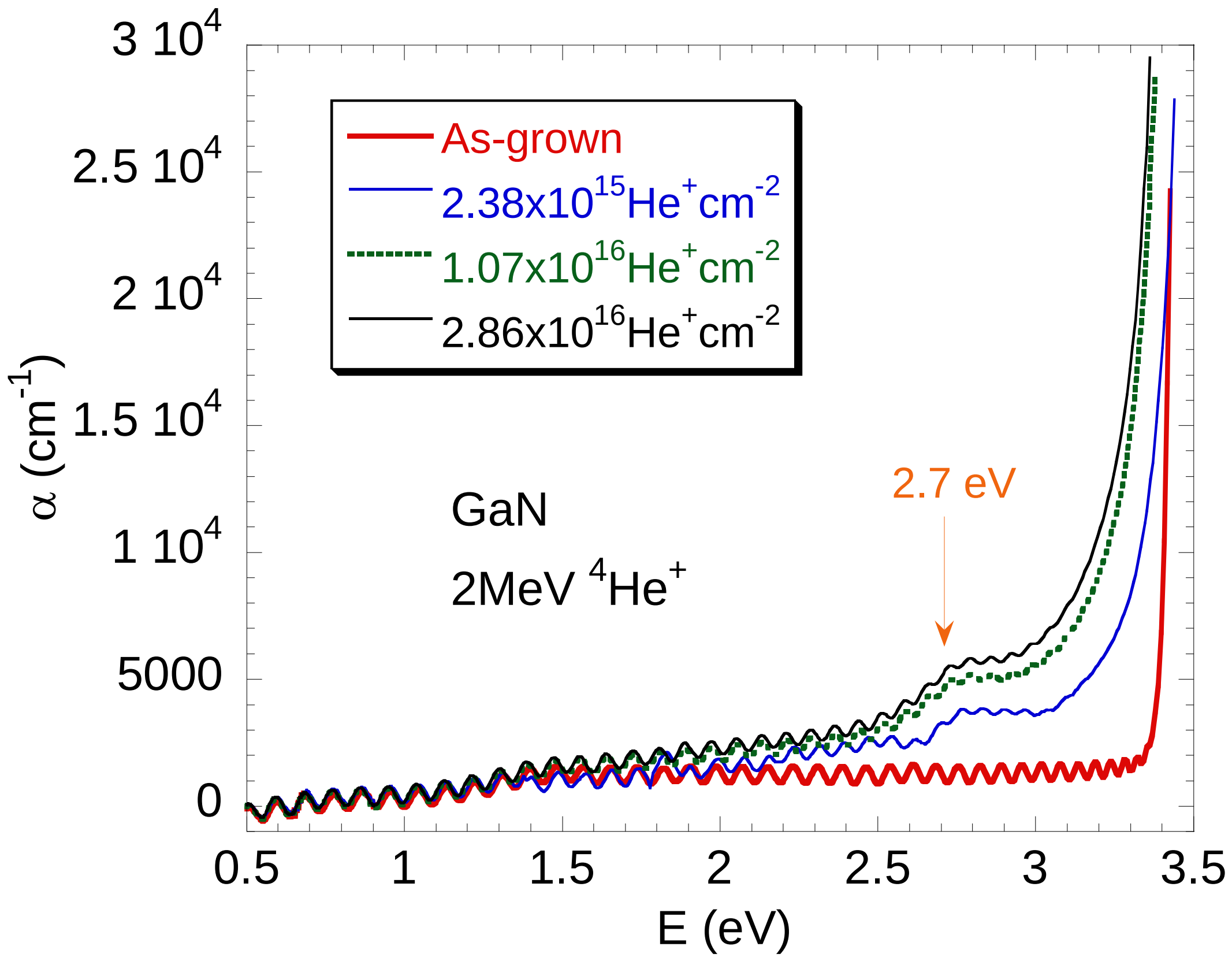

Fig. 2 of 3.

Li *et al*.

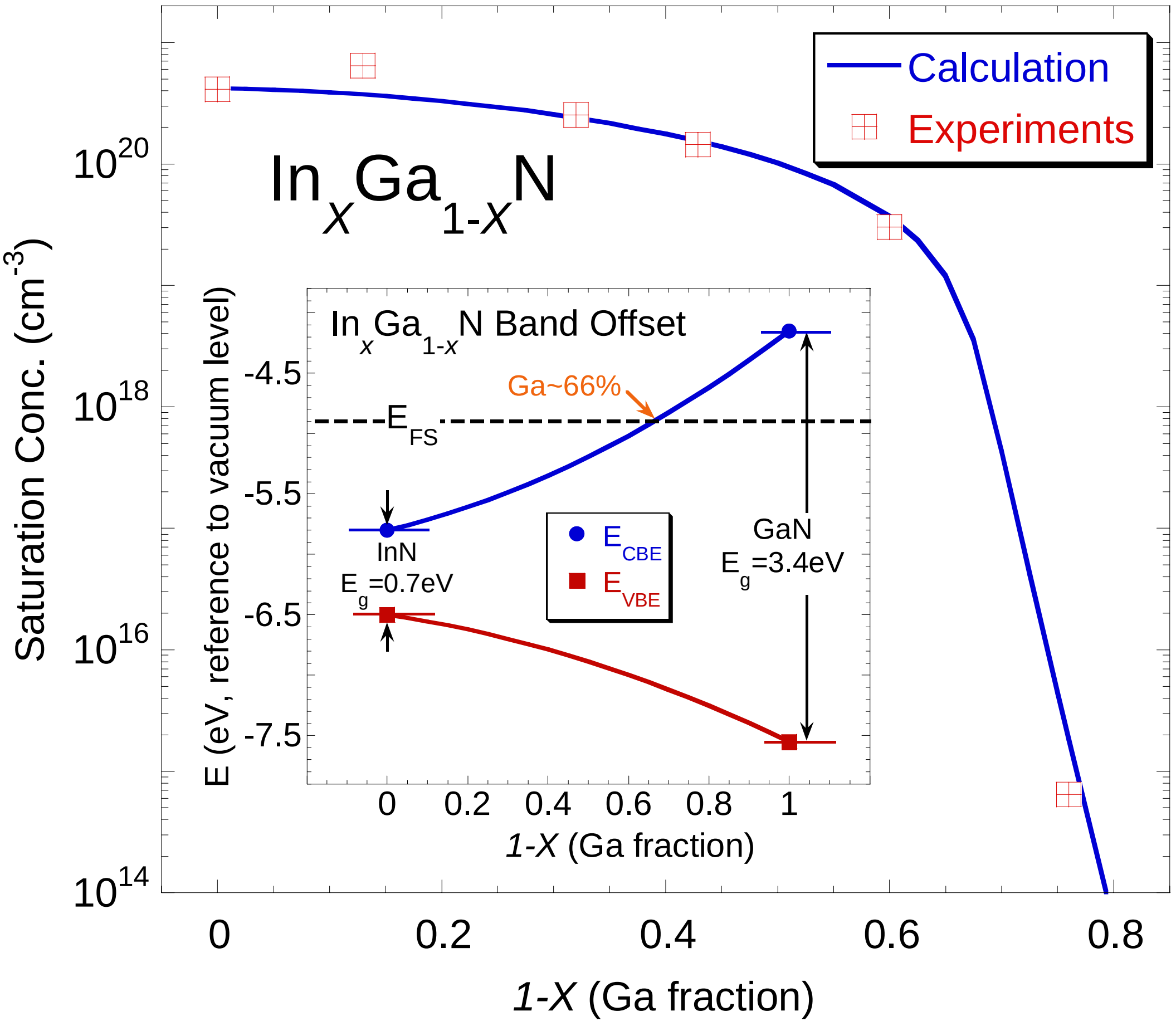

Fig. 3 of 3

Li *et al.*